\documentclass[aps,pra, showpacs,amssymb,superscriptaddress,nofootinbib, onecolumn]{revtex4-2} 

\usepackage[utf8]{inputenc}
\usepackage{xcolor}
\usepackage{graphicx} 
\usepackage{epsfig}
\usepackage{epstopdf}
\usepackage{braket}
\usepackage[normalem]{ulem}
\usepackage[colorlinks]{hyperref}
\usepackage{siunitx}
\usepackage{soul}
\usepackage{mathtools}
\usepackage{float}
\begin{document}



\title{One-photon-interference quantum secure direct communication}

\author{Xiang-Jie Li}
\affiliation{State Key Laboratory of Low-dimensional Quantum Physics and Department of Physics, Tsinghua University, Beijing 100084, China}

\author{Min Wang}
\affiliation{Beijing Academy of Quantum Information Sciences, Beijing 100193, China}

\author{Xing-Bo Pan}
\affiliation{State Key Laboratory of Low-dimensional Quantum Physics and Department of Physics, Tsinghua University, Beijing 100084, China}

\author{Yun-Rong Zhang}
\affiliation{State Key Laboratory of Low-dimensional Quantum Physics and Department of Physics, Tsinghua University, Beijing 100084, China}

\author{Gui-Lu Long}%
\email{gllong@mail.tsinghua.edu.cn}
\affiliation{State Key Laboratory of Low-dimensional Quantum Physics and Department of Physics, Tsinghua University, Beijing 100084, China}
\affiliation{Beijing Academy of Quantum Information Sciences, Beijing 100193, China}
\affiliation{Frontier Science Center for Quantum Information, Beijing 100084, China}
\affiliation{Beijing National Research Center for Information Science and Technology, Beijing 100084, China}

\date{\today}

\begin{abstract}
  Quantum secure direct communication (QSDC) is a quantum communication paradigm that transmits confidential messages directly using quantum states. Measurement-device-independent (MDI) QSDC protocols can eliminate the security loopholes associated with measurement devices. To enhance the practicality and performance of MDI-QSDC protocols, we propose a one-photon-interference MDI QSDC (OPI-QSDC) protocol which transcends the need for quantum memory, ideal single-photon sources, or entangled light sources. The security of our OPI-QSDC protocol has also been analyzed using quantum wiretap channel theory. Furthermore, our protocol could double the distance of usual prepare-and-measure protocols, since quantum states sending from adjacent nodes are connected with single-photon interference, which demonstrates its potential to extend the communication distance for point-to-point QSDC.
\end{abstract}

\maketitle
\clearpage

\section{\label{sec:level1}Introduction\protect\\ }
Quantum communication uses physical principles to ensure the security of communication. Bennett and Brassard proposed quantum key distribution (QKD) protocol in 1984 (BB84)~\cite{BB84} which provides a secure way for key agreement. Long and Liu proposed quantum secure direct communication (QSDC)~\cite{LL00} in 2000, which provides secure and reliable communication in a channel with both noise and eavesdropping~\cite{LL00}. 

Rapid inevitable developments have been made in the fields of QKD and QSDC~\cite{BB84,LL00,DL04,wang2005high,wang2005multi,deng2004bidirectional,zhang2020design,kish2020feasibility,pan2020simultaneous,cui2019measurement,kwek2021chip,guo2021toward,sun2021deterministic}. QSDC transmits information directly through a quantum channel, using entanglement~\cite{LL00,TwoStep03,wang2005high,wang2005multi,twoQSDD,pan2020single,cao2021continuous} or single photons~\cite{DL04,pan2020single}. The security of these protocols has been completed~\cite{qi2019,wu2019security,wu2022quantum} based on the quantum wiretap channel theory~\cite{wyner1975wire,Devetak2005,cai2004quantum}. Due to imperfections in measurement devices, practical systems possess security loopholes~\cite{qi2005,Makarov*2005,makarov2009}. In order to fix these loopholes, measurement-device-independent QSDC (MDI-QSDC)~\cite{niu2018measurement,zhou2020measurement,niu2020security} protocols have been developed. In 2021, Long \emph{et al.} proposed a simple and powerful method to increase channel capacity using masking (INCUM)~\cite{long2021drastic}, which increases the channel capacity and communication distance of QSDC. There have been several successful demonstrations of QSDC in fiber~\cite{qi2019,hu16experimental,zhu17experimental} and in free space~\cite{pan2020experimental,zhang17experimental}. In 2020, a secure-classical repeater QSDC network have been proposed and experimentally demonstrated~\cite{long2020globecom,long2022evolutionary}. A 15-user QSDC network with direct links has been reported~\cite{qi202115}. In 2022, Zhang \emph{et al.} report a QSDC system with mixed coding of phase and time-bin states and achieves a communication distance of 102.2 km in fiber, setting up a new world record~\cite{zhang2022realization}.

Traditional QSDC protocols~\cite{LL00,TwoStep03,DL04,niu2018measurement,zhou2020measurement} require two-way transmission, which leads to double loss of signal in the channel and also limits the transmission distance by half compared to one-way protocols in principle. Remarkably, with the help of the quantum-memory-free (QMF) technique, one-way QSDC protocols have been proposed~\cite{wen2007reusable,deng2019repeatable} and QMF-QSDC with sophisticated coding has also been designed~\cite{sun2018design,sun2020qmf}. In addition, there are some one-way QSDC protocols using hyperentanglement~\cite{zhou2023device,sheng2022one,zhou2022one,ying2022measurement}.

However, the MDI-QSDC~\cite{niu2018measurement,zhou2020measurement,niu2020security} protocols face several practical limitations, including reliance on immature quantum memory, the assumption of ideal entangled or single-photon light sources, limited transmission distances and low secrecy rates. Single-photon-memory MDI-QSDC~\cite{li2023single} utilizes QMF technique to eliminate the dependency of MDI-QSDC protocol on high-performance quantum memory. To further address other issues, we propose a one-photon-interference MDI QSDC (OPI-QSDC) protocol in this paper. Our OPI-QSDC protocol is a new one-way MDI-QSDC protocol that operates without relying on quantum memory. Moreover, we harnesses single-photon interference, as utilized in twin-field QKD protocols~\cite{lucamarini2018overcoming, ma2018phase, wang2018twin, curty2019simple, cui2019twin} that have broken the repeaterless quantum communications bound known as the Pirandola-Laurenza-Ottaviani-Banchi (PLOB) bound~\cite{pirandola2017fundamental}. In the OPI-QSDC protocol, Alice and Bob are able to achieve single-photon interference at the intermediate node Charlie by utilizing phase-locking techniques~\cite{wang2022twin}, when using weak coherent light sources. This doubles the transmission distance compared to other one-way QSDC protocols~\cite{deng2019repeatable}. We analyze the security of our protocol using quantum wiretap channel theory and simulate its performance, demonstrating its ability to break the PLOB bound.

The remainder of this paper is organized as follows. In Sec.~\ref{sec:level2} we describe the detailed steps of the proposed OPI-QSDC protocol, while in Sec.~\ref{sec:level3} we analyze its security. Then in Sec.~\ref{sec:level4} we present our numerical analysis of performance. Finally, we give a conclusion in Sec.~\ref{sec:level5}. 

\begin{figure*}[t]
  \centering
   \includegraphics[width=\textwidth]{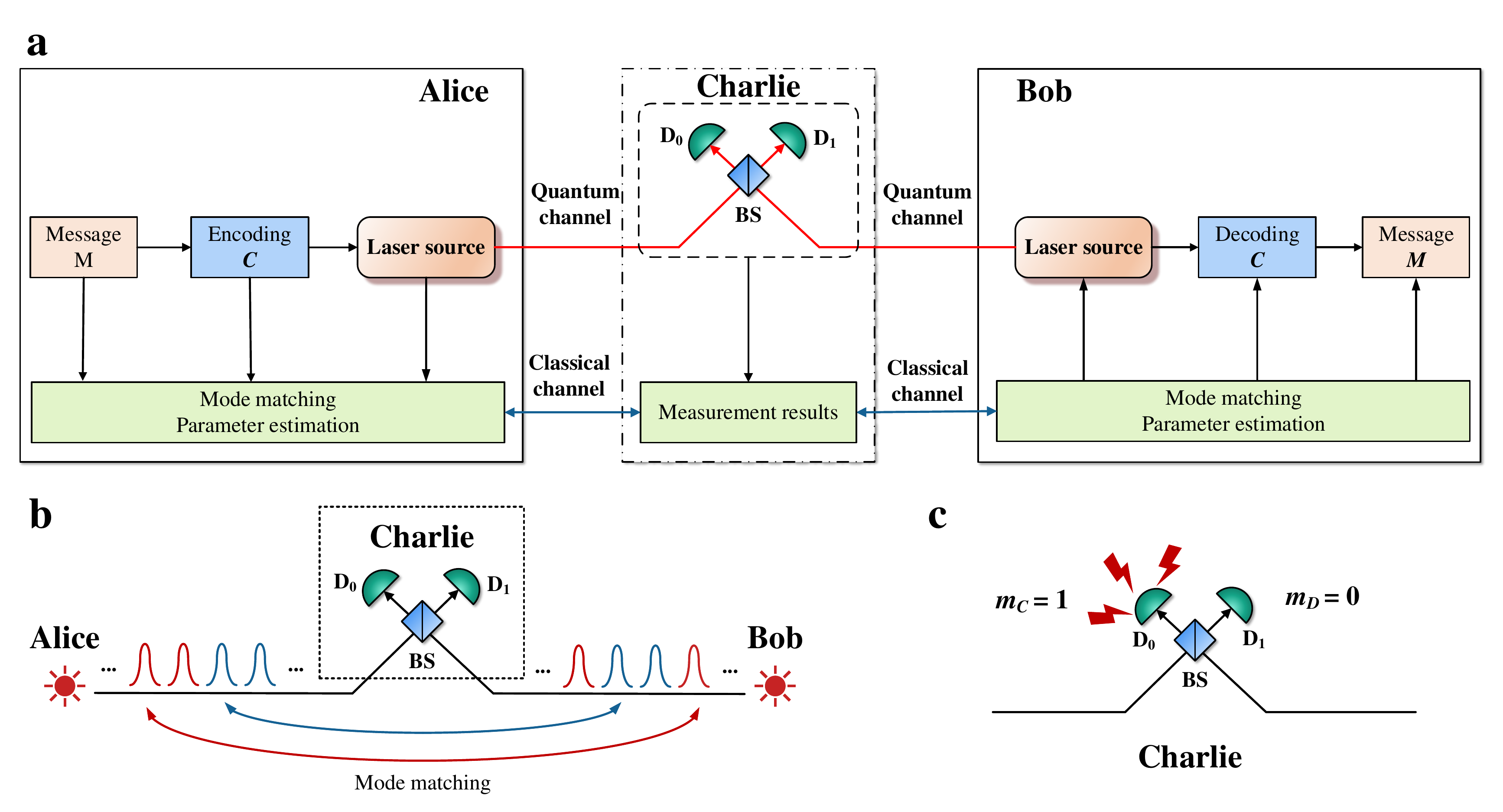}
   \caption{\label{fig:diagram}\textbf{Schematic diagram of OPI-QSDC protocol and the process of mode matching and measurement.} BS, 50:50 beam splitter; $\rm D_0$, $\rm D_1$, single-photon detector. \textbf{a.} In the OPI-QSDC protocol, we assume that the laser source and coding process of Alice and Bob will not be attacked by Eve, which is marked with a solid box. The untrusted Charlie uses a single-photon interferometer to measure the signals sent by Alice and Bob, but the measurement is completely controlled by Eve, which is marked with a dashed box. The three communication parties also need to use the open classic channel marked with the green box to exchange information. We assume that the information in this channel published by Alice and Bob will not be tampered with by Eve.  \textbf{b.} Red and blue pulses represent coding mode and multi-intensity mode, respectively. The same mode sent by Alice and Bob will be able to achieve a mode matching, and deterministic transmission of information. \textbf{c.} we take the case where only detector $\rm D_0$ clicks as an example. At this time, Charlie publishes that $m_C$=1, $m_D$=0.
   }
\end{figure*}

\section{\label{sec:level2}Details of protocol\protect\\ 
}
In this section, we propose our OPI-QSDC protocol. As illustrated in Fig.~\ref{fig:diagram}a, we suppose that Alice and Bob utilize phase locking techniques to lock the frequency and global phase of their lasers. They then simultaneously send light pulses to Charlie, an untrusted third party right in the middle between them. We use $M\in\{0,1\}^m$ to represent the message that Alice wants to transmit to Bob, and $C\in\{0,1\}^c$ represents the ciphertext. In particular, The detailed steps of our protocol are as follows.

\textbf{Step 1: encoding.} Alice encodes the message $M$ to be transmitted to form the codeword $C$. Encoding operations use forward error correction codes, secure codes~\cite{sun2020qmf} and INCUM~\cite{long2021drastic}. Details of encoding process can be found in the Appendix~\ref{app:encoding}.

\textbf{Step 2: mode preparation.} Alice and Bob randomly select the coding mode with a probability of $1-p$ and the multi-intensity mode with a probability of $p$, where $0<p<<1$. The quantum states sent in each mode are as follow.

\setlength{\parindent}{2em}Coding mode: a weak coherent state $|\alpha \rangle$ for logical $0$ or a weak coherent state $|-\alpha \rangle$ for logical $1$, where $|\alpha|^2=u$ is the intensity of the coherent state. In this mode, Alice selects the coherent state to be sent based on the encoding result. For example, if the encoding result is $0$, Alice sends a $|\alpha \rangle$ state; otherwise, she sends a $|-\alpha \rangle$ state. Bob randomly chooses to send these two states.

\setlength{\parindent}{2em}Multi-intensity mode: three different intensity and phase-randomized weak coherent states $\hat{\rho}_{\beta_a} =|\beta_a e^{i\phi_a} \rangle$ and $\hat{\rho}_{\beta_b}=|\beta_b e^{i\phi_b} \rangle$, where $\beta_a$ and $\beta_b$ are randomly choosen from the set of $\{\sqrt{\nu_1}, \sqrt{\nu_2},0\}$, while $u>>\nu_1>\nu_2>0$, and $\phi_a$, $\phi_b$ are randomly choosen from $[0,2\pi)$.

\textbf{Step 3: measurement.} Charlie measures the pulses sent by Alice and Bob using single-photon interferometer, and publishes measurement results. Let $m_C$ and $m_D$ denote the measurement outcome of $\rm D_0$ and $\rm D_1$, where value "0" indicates a no-click event and  value "1" indicates a click event. As shown in Fig.~\ref{fig:diagram}c. Alice and Bob discard the no-click events and two-click events, and retain the one-click events, namely $m_C\oplus m_D=1$.

\textbf{Step 4: mode matching.} After all measurements are completed, Alice and Bob publish the modes information. They retain the measurement results of the same modes, and discard the measurement results of different modes. As shown in Fig.~\ref{fig:diagram}b. Note that there is a probability of $2p(1-p)$ for a mode mismatch, resulting in the loss of information transmitted by Alice. However, Alice and Bob can utilize error correcting codes in \textbf{Step 1} to recover the lost information. If they both send the multi-intensity mode, they publish the intensity and phase of the weak coherent state. Then they retain pulses with $\beta_a=\beta_b$ and $|\phi_a-\phi_b|=0$ or $\pi$.

\textbf{Step 5: parameter estimation.} Alice and Bob randomly publish the bit values in some coding modes to estimate the quantum bit error rate (QBER), and then use different intensity valuses in multi-intensity modes to estimate the amount of information leakage. Based on these results of parameter estimation, they proceed to step 6.

\textbf{Step 6: decoding.} Bob decodes the message $M$ from the codeword $C$. Details of decoding process can be found in the Appendix~\ref{app:encoding}.

\section{\label{sec:level3}Security analysis\protect\\ }

In order to complete our security proof, we introduce an equivalent entanglement-based OPI-QSDC protocol, as detailed in Appendix~\ref{app:Security}. In this protocol, we transform the laser source into an entanglement-photon source that can be analyzed conveniently. This way, the security of entanglement-based protocol will imply the security of OPI-QSDC.

According to quantum wiretap channel theory~\cite{wyner1975wire,thangaraj2007applications,tyagi2015universal,wu2019security,niu2020security,wu2022quantum}, There is a secrecy capacity $C_s=C_M-C_W$ that enables us to reliably and securely transmit the message to recipients by using a forward encoding with a coding rate $R$ lower than it, where $C_M$ and $C_W$ are the main channel's capacity and wiretap channel's capacity, respectively. We discard the case where there is no detector click and both detectors click, then first consider the case where only detector $\rm D_0$ clicks. In this case, the achievable secrecy rate is
\begin{equation}
R^C=I^C(A:B)-I^C(A:E),
\end{equation}
where $I^C(A:B)$ is the mutual information of Alice and Bob when only detector $\rm D_0$ clicks, while $I^C(A:E)$ is the mutual information of Alice and Eve when only detector $\rm D_0$ clicks. We assume the channel of Alice and Bob is a symmetric channel, thus 
\begin{equation}
  I^C(A:B)=1-h(e),
\end{equation}
where $h(x)$ is the binary entropy function, i.e. $h(x)=-x\log_2(x)-(1-x)\log_2(1-x)$, and $e$ is quantum bit error rate (QBER). In our protocol, we use X basis to transmit information, so $e=E_u^{X,C}$, where $E_u^{X,C}$ is the X-basis error rate when only detector $\rm D_0$ clicks. 

To calculate $I^C(A:E)$, we can analyze the process of eavesdropping and then use the results of the parameter estimation. See Appendix~\ref{app:Security} for details of the derivation. The upper bound on $I^C(A:E)$ is
\begin{equation}
  \begin{aligned}
  I^C(A:E)&\leq h(E_u^{Z,C})
  &=h[\frac{1}{Q^C_u} &(\sum_{n = 0}^{\infty} \sqrt{|C_{2n}|^2 Y^C_{2n}})^2]
  \end{aligned}
\end{equation}
where $Q^C_u$ is the total gain, i.e., the conditional probability of only detector $\rm D_0$ clicks when Alice and Bob send pulses with the intensity of $u$, while $|C_{2n}|^2$ is the probability when there are $2n$ photons in the channel. $Y^C_{2n}$ is the yield of the $2n$-photon state, i.e., the conditional probability of only detector $\rm D_0$ clicks when there are $2n$ photons in the channel. Hence the achievable rate is
\begin{equation}
  \begin{aligned}
   R^C=&I^C(A:B)-I^C(A:E)\\
   \geq&  q\cdot Q^C_u\cdot [1-fh(E^{X,C}_u)-h(E^{Z,C}_u)],
  \end{aligned}
\end{equation}
where $q=1-2p(1-p)$ is the mode matching rate, and $f\geq 1$ is an inefficiency function for forward coding. We skip the discussion for only $\rm D_1$ clicks, however, it holds in a similar manner, which is
\begin{equation}
  \begin{aligned}
   R^D\geq q\cdot Q^D_u\cdot [1-fh(E^{X,D}_u)-h(E^{Z,D}_u)].
  \end{aligned}
\end{equation}
Finally, the total secrecy rate formula is given by
\begin{equation}
  \begin{aligned}
   R=& \max\{R^C,0\}+\max\{R^D,0\}.
  \end{aligned}
\end{equation}

\section{\label{sec:level4}Performance analysis\protect\\ }
\subsection{Comparison with other QSDC protocols}
We performed numerical simulations for characterizing the performance of the proposed OPI-QSDC and other QSDC protocols~\cite{niu2018measurement,DL04}. The key parameter settings for our simulations are shown in Table \ref{Tab:parameter}~\cite{ma2018phase,zhang2022realization}.
\begin{table}[htbp]
  \centering
  \caption{\label{Tab:parameter}Key parameter settings of simulation.}
  \begin{tabular}{ccc}
  \hline 
  \hline
Parameter & Value              & Description  \\ \hline 
$\zeta$  & 0.2 dB/km          & the attenuation coefficient \\
$\eta_d$  & 15\%             & the efficiency of detectors \\
$p_d$     & $8\times10^{-8}$   & the probability of dark count \\
$\delta$     & 1.5\%               & the misalignment probability  \\
$f$       & 1.2                & the inefficiency function for forward coding\\
$u$       & 0.046                & the light intensity\\
\hline
\hline
\end{tabular}
\end{table}

\begin{figure}[h]
  \centering
   \includegraphics[width=\columnwidth]{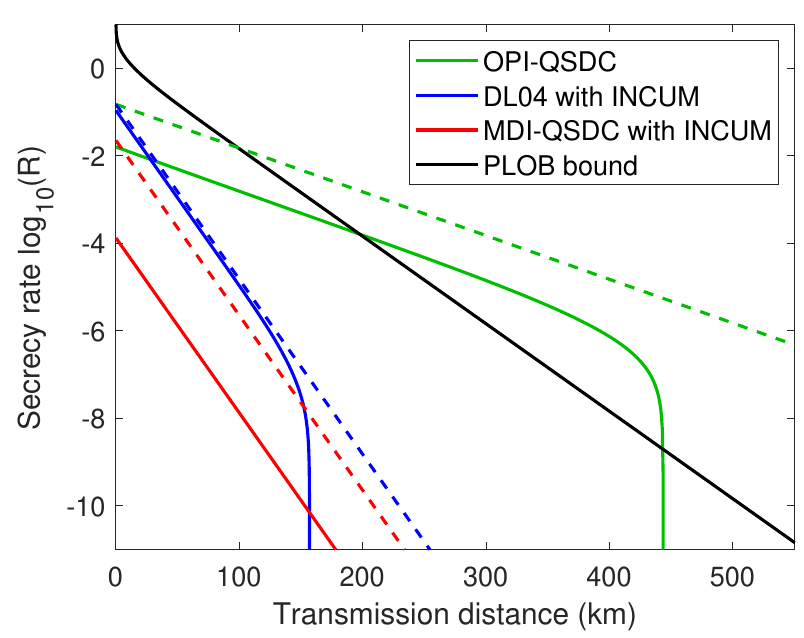}
   \caption{\label{fig:comparison}\textbf{Secrecy rate $log_{10}(R)$ versus the transmission distance of OPI-QSDC and other QSDC protocols.}  The black line is the PLOB bound. The green, blue, and red line represent OPI-QSDC, DL04 and MDI-QSDC protocol, respectively. The solid (dashed) lines indicate the parameterized (idealized) protocol performance. For the specific parameters, the rate of OPI-QSDC surpasses the PLOB bound when $d>$ 228 km. The longest trasmission distance of our protocol is able to achieve 440 km.
   }
\end{figure}

As shown in Fig.~\ref{fig:comparison}, the OPI-QSDC protocol is able to exceed the PLOB bound when $d>$ 228 km with practical parameters. The Appendix \ref{app:Performance} contains derivation details of simulation formulas. We also simulate the performance of MDI-QSDC~\cite{niu2018measurement} and DL04~\cite{DL04} protocol with the method of INCUM~\cite{long2021drastic}. The MDI-QSDC and DL04 protocols are two-way protocols which suffer double channel loss over a certain transmission distance. To be more precise, the MDI-QSDC protocol detects twice to complete the transmission of information, while DL04 protocol detects only once. Therefore, the explicit secrecy rate equations used to draw the idealized protocol performance in Fig.~\ref{fig:comparison} are: 
\begin{equation}
  \begin{aligned}
    &R_{PLOB}^{ideal}=-\log_2(1-\eta_c),\\
    &R_{OPI-QSDC}^{ideal}=\eta_d\sqrt{\eta_c},\\
    &R_{DL04}^{ideal}=\eta_d\eta_c^2,\\
    &R_{MDI-QSDC}^{ideal}=(\eta_d\eta_c)^2,
  \end{aligned}
\end{equation}
where $\eta_d$, $\eta_c$ are the detection efficiencies and the channel losses function, respectively. The DL04 protocol's beginning rates will be higher than the OPI-QSDC, since we did not account for the effects of real light sources on its performance when we parameterized the protocols. The Appendix \ref{app:DL04andMDI} contains further simulation details of the DL04 and MDI-QSDC protocols.

\subsection{Effect of light intensity and dark counts}
The effects of light intensity $u$ and dark counts $p_d$ on the OPI-QSDC protocol have been explored separately. The data in Table \ref{Tab:parameter} are utilized to determine the parameters of the numerical simulations in this subsection. 

\begin{figure}
  \centering
   \includegraphics[width=\columnwidth]{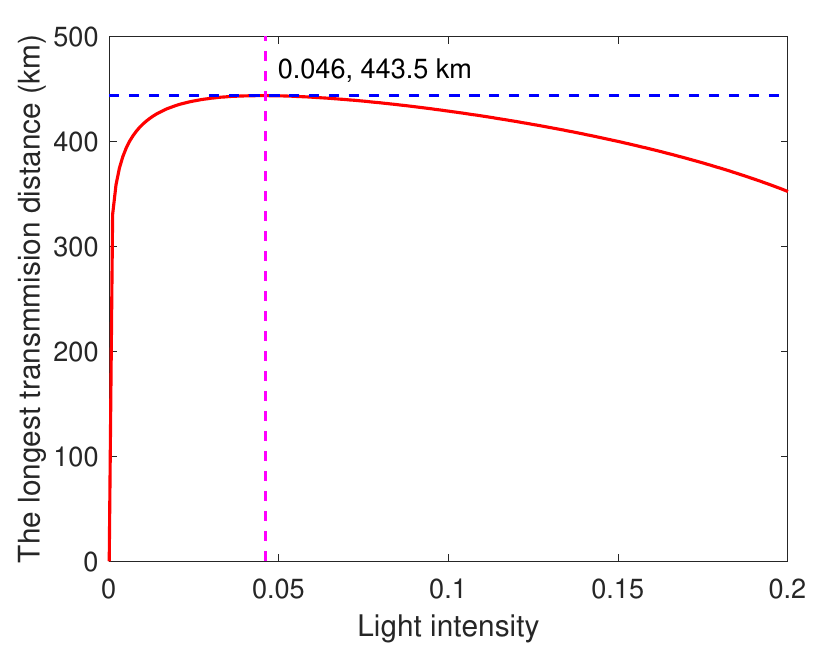}
   \caption{\label{fig:light}\textbf{The longest transmission distance versus the light intensity $u$.} The red-solid line indicates the longest distance that can be transmitted by OPI-QSDC protocol under different light intensities. The blue-dashed line $d$=443.5 km and the pink-dashed line $u$=0.046 mark the maximum distance and corresponding light intensity that can be transmitted by our protocol.
   }
\end{figure}

In terms of light intensity, there are two aspects to consider. Firstly, as the light intensity increases, the mean photon number in the channel also increases, leading to a corresponding increase in the gain $Q_u$. Secondly, the even-photon-number component in the channel becomes more prominent, resulting in a higher phase-error rate $E_u^Z$. As illustrated in Fig.~\ref{fig:light}, the OPI-QSDC protocol demonstrates its optimal performance when the light intensity is set to $u = 0.046$. With these parameters, our OPI-QSDC protocol achieves a maximum transmission distance of 443.5 km. This finding underscores the significance of employing a relatively weaker coherent state light source to enhance the protocol's performance in practical applications. However, it's worth noting that an excessively weak light source may also lead to a low transmission distance.

\begin{figure}
  \centering
   \includegraphics[width=\columnwidth]{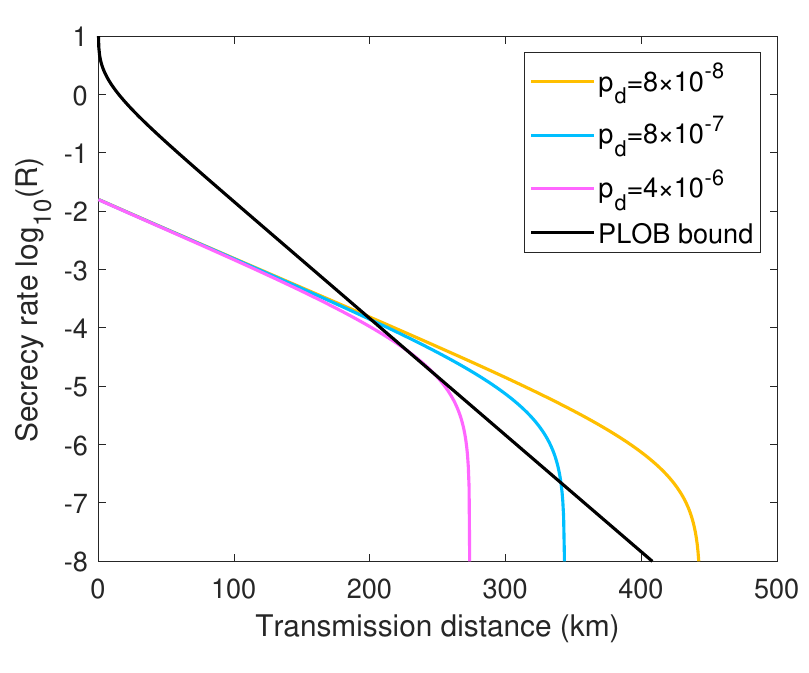}
   \caption{\label{fig:pd}\textbf{Secrecy rate $log_{10}(R)$ versus the transmission distance parameterized by different dark counts $p_d$.} The black line is the PLOB bound. The yellow, blue and pink lines indicate the secrecy rate R at dark counts $p_d=8\times 10^{-8}$, $8\times 10^{-7}$ and $4\times 10^{-6}$ changing with transmission, respectively.
   }
\end{figure}

For the dark counts, as the transmission distance increases, the signal light will attenuate, when the rate of the dark count is able to compare to the signal light, both QBER and phase-error rate increase dramatically, thus limiting the longest transmission distance of the signal. As shown in Fig.~\ref{fig:pd}, our protocol can still reach the PLOB bound when the dark count rate reaches $4\times 10^{-6}$.

\section{\label{sec:level5}Conclusions\protect\\ }
In summary, our OPI-QSDC protocol has been proposed, and its security has been analyzed when using practical coherent light sources. The performance analysis of our protocol shows that, compared to the DL04 and MDI-QSDC protocols, it has a longer transmission distance which could achieve 443.5 km with a light intensity of 0.046 when using non-ultra-low loss optical fiber. Additionally, it can surpass the PLOB bound when the transmission distance exceeds 228 km.

In the future, the OPI-QSDC protocol has the potential to be put into practical application and may find applications in the global quantum Internet.

\begin{acknowledgements}
  Min Wang acknowledges the Young Elite Scientists Sponsorship Program by China Association for Science and Technology (2022QNRC001). This work is supported by the National Natural Science Foundation of China under Grants No. 11974205 and No. 12205011, the Key R\&D Program of Guangdong province (2018B030325002), Beijing Advanced Innovation Center for Future Chip (ICFC), Tsinghua University Initiative Scientific Research Program.
\end{acknowledgements}

\clearpage

\appendix
\section{Details of encoding and decoding}
\label{app:encoding}
\begin{figure}
  \centering
   \includegraphics[width=\columnwidth]{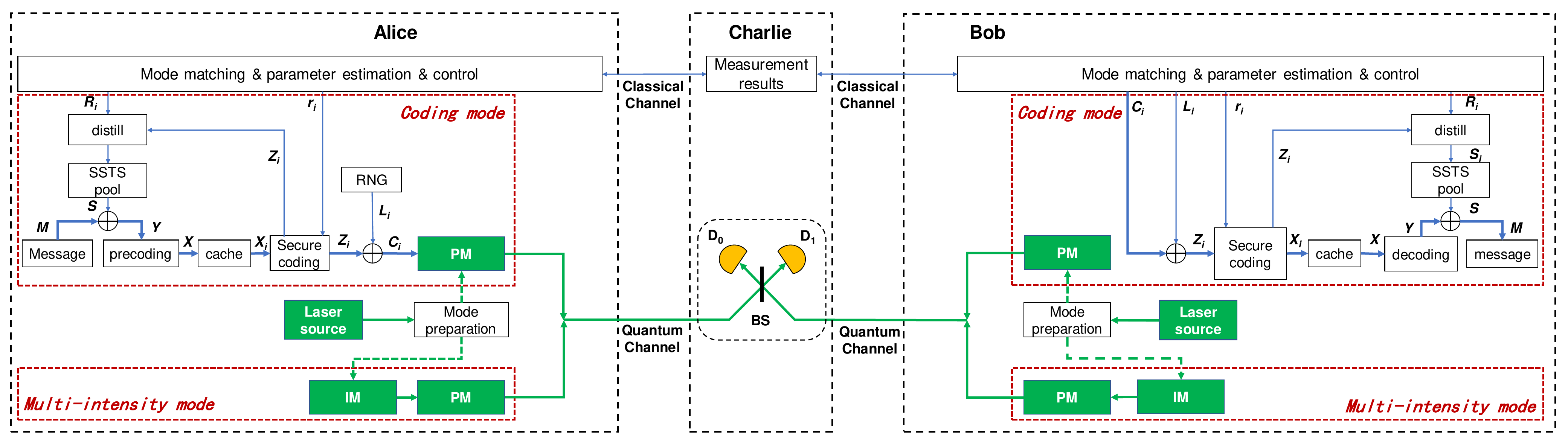}
   \caption{\label{fig:encoding}\textbf{The schematic diagram of the OPI-QSDC protocol in $i$-th frame with detailed encoding and decoding process.} SSTS: shared secure transmission sequence. IM: intensity modulator. PM: phase modulator. RNG: random number generator.
   }
\end{figure}
To ensure the security of QSDC, it is essential to establish the channel's security before loading information. To achieve this, in the OPI-QSDC protocol, the message to be transmitted is divided into multiple frames, with each frame containing several bits. The security of each frame is guaranteed by using the estimation results of the previous frame parameters. The first frame is used for transmitting random numbers. The schematic diagram of the OPI-QSDC protocol in $i$-th frame, including the complete encoding and decoding process, is illustrated in Fig.~\ref{fig:encoding}. We use the dynamic joint encoding proposed in \cite{sun2020qmf}. Here, $M \in \{0,1\}^m$ represents the plaintext message that Alice wants to send to Bob. $S \in \{0,1\}^m$ is a shared secure transmission sequence (SSTS) extracted from the SSTS pool, which is used for encoding. $Y$ is the encrypted message, where $Y = M \oplus  S$. $ X \in \{0,1\}^k$ is a $k$-length $(k, kR_p)$ LDPC codeword obtained by performing LDPC precoding on $Y$, resulting in an output rate of $kR_p = m$. Sequence $X_i \in \{0,1\}^{k_i}$ is a part of X or a randomly generated bit sequence, serving as the input to the secure encoding module in the $i$-th frame. $r_i$ is the rate of secure encoding in the $i$-th frame. $Z_i \in \{0,1\}^{n_{ci}}$ is the encoded text generated from sequence $X_i$ with a length of $n_{ci}$. $L_i \in \{0,1\}^{n_{ci}}$ represents locally generated random numbers used for INCUM encryption,$ C_i = Z_i \oplus L_i$. After transmitting $C_i$, Alice and Bob can infer the mutual information between Alice and Bob $I(A:B)_i$, and Alice and Eve $I(A:E)_i$. Based on these, they can obtain the secrecy rate $R_i$ for the $i$-th frame. The above parameters must satisfy the condition of 
\begin{equation}
  \label{equ:CMCW}
  \begin{aligned}
\frac{k_i}{n_{ci}}&\leq r_i-I(A:E)_{i-1},\\
r_i&<I(A:B)_{i-1},
  \end{aligned}
  \end{equation}
where $I(A:E)_{i-1}=q\cdot Q^{i-1}_u\cdot h(E_u^{Z,i-1})$ and $I(A:B)_{i-1}=q\cdot Q^{i-1}_u\cdot [1-f\cdot h(E_u^{X,i-1})]$, based on Section~\ref{sec:level3}. These two conditions ensure the security and reliability of information transmission, respectively.

The detailed steps of $i$-th frame are as follows.

\textbf{Step 1: encoding.} Alice encodes the message $M$ to be transmitted as follows.

(1). Alice uses SSTS $S$ to encrypt plaintext message $M$ and obtains $Y$, where $Y = M \oplus  K$. Note that in the first round of communication, the SSTS pool is empty. At this point, Alice can generate $X_i$ using random numbers and transmit it to Bob. Then, they can estimate $I(A:B)_1$ and $I(A:E)_1$, and extract the same SSTS that satisfies equation~\ref{equ:CMCW}. They can then add it to the SSTS pool. From this point onwards, there will be sufficient SSTS available for further communication.

(2). Alice encodes $Y$ into X and then places it in the cache.

\textbf{Step 2: mode preparation.} Alice and Bob randomly select the coding mode with a probability of $1-p$ and the multi-intensity mode with a probability of $p$, where $0<p<<1$. The quantum states sent in each mode are as follow.

\setlength{\parindent}{1em}Coding mode: If Alice chooses this mode, she continues the encoding process:

\setlength{\parindent}{2em}(1). Alice selects $k_i$ bits from the cache to perform secure encoding, resulting in $Z_i$.

\setlength{\parindent}{2em}(2). Alice generates a local random bit string $L_i$ and encrypts $Z_i$ using it to obtain $C_i$, where $C_i = Z_i \oplus L_i$.

\setlength{\parindent}{2em}(3). Alice maps each bit of $C_i$ into a quantum state. She prepares the quantum state $|\alpha \rangle$ to represent logical bit 0, and quantum state $|-\alpha \rangle$ to represent logical bit 1.

\setlength{\parindent}{2em}(4). Alice sends the quantum states to Charlie.

If Bob chooses this mode, he randomly chooses to send state $|\alpha \rangle$ or $|-\alpha \rangle$.

\setlength{\parindent}{1em}Multi-intensity mode: three different intensity and phase-randomized weak coherent states $\hat{\rho}_{\beta_a} =|\beta_a e^{i\phi_a} \rangle$ and $\hat{\rho}_{\beta_b}=|\beta_b e^{i\phi_b} \rangle$, where $\beta_a$ and $\beta_b$ are randomly choosen from the set of $\{\sqrt{\nu_1} , \sqrt{\nu_2},0\}$.

\textbf{Step 3: measurement.} Charlie measures the pulses sent by Alice and Bob using single-photon interferometer, and publishes measurement results. Let $m_C$ and $m_D$ denote the measurement outcome of $\rm D_0$ and $\rm D_1$, where value "0" indicates a no-click event and  value "1" indicates a click event. Alice and Bob discard the no-click events and two-click events, and retain the one-click events, namely $m_C\oplus m_D=1$.

The \textbf{step 2} and \textbf{step 3} are repeated for several rounds until the transmission of $C_i$ is complete.

\textbf{Step 4: mode matching.} After all measurements are completed, Alice and Bob publish the modes information. They retain the measurement results of the same modes, and discard the measurement results of different modes. If they both send the multi-intensity mode, they publish the intensity and phase of the weak coherent state. Then they retain pulses with $\beta_a=\beta_b$ and $|\phi_a-\phi_b|=0$ or $\pi$.

\textbf{Step 5: parameter estimation.} After completing the above steps, Alice and Bob can obtain the gain $Q_u^{i}$ for the current round of communication. They then randomly publish some of the bit values from the coding mode to estimate the quantum bit error rate $E_u^{X, i}$ and use different intensity valuses in multi-intensity modes to estimate the phase-error rate $E_u^{Z, i}$. Based on this information, they can calculate the mutual information between Alice and Bob $I(A:B)_{i}$, Alice and Eve $I(A:E)_i$, and the secrecy rate $R_{i}$. If $I(A:B)_{i}$ and $I(A:E)_{i}$ satisfy equation~\ref{equ:CMCW}, they can extract the same SSTS from $C_i $ and add it to the SSTS pool for the next round of communication.

The \textbf{Step 2} to \textbf{Step 5} are repeated for several rounds until the transmission of X is complete.

\textbf{Step 6: decoding.} Bob publishes the positions of his measurements which have valid results and Alice publishes the value of the local random
bit $L_i$ at these positions to help Bob unmask the codeword $Z_i$. Then Bob uses the same $S$ to decrypt $Y$ and obtain the plaintext message $M$, where $M = Y \oplus S$.

If the SSTS pool is depleted during the transmission process, Alice and Bob will need to transmit $X_i$, which is composed of random numbers and satisfies equation~\ref{equ:CMCW},  to extract the same $S$. This ensures that they have sufficient SSTS to continue the communication securely.

\section{Details of security analysis}
\label{app:Security}
\begin{figure}[h]
    \centering
     \includegraphics[width=350pt]{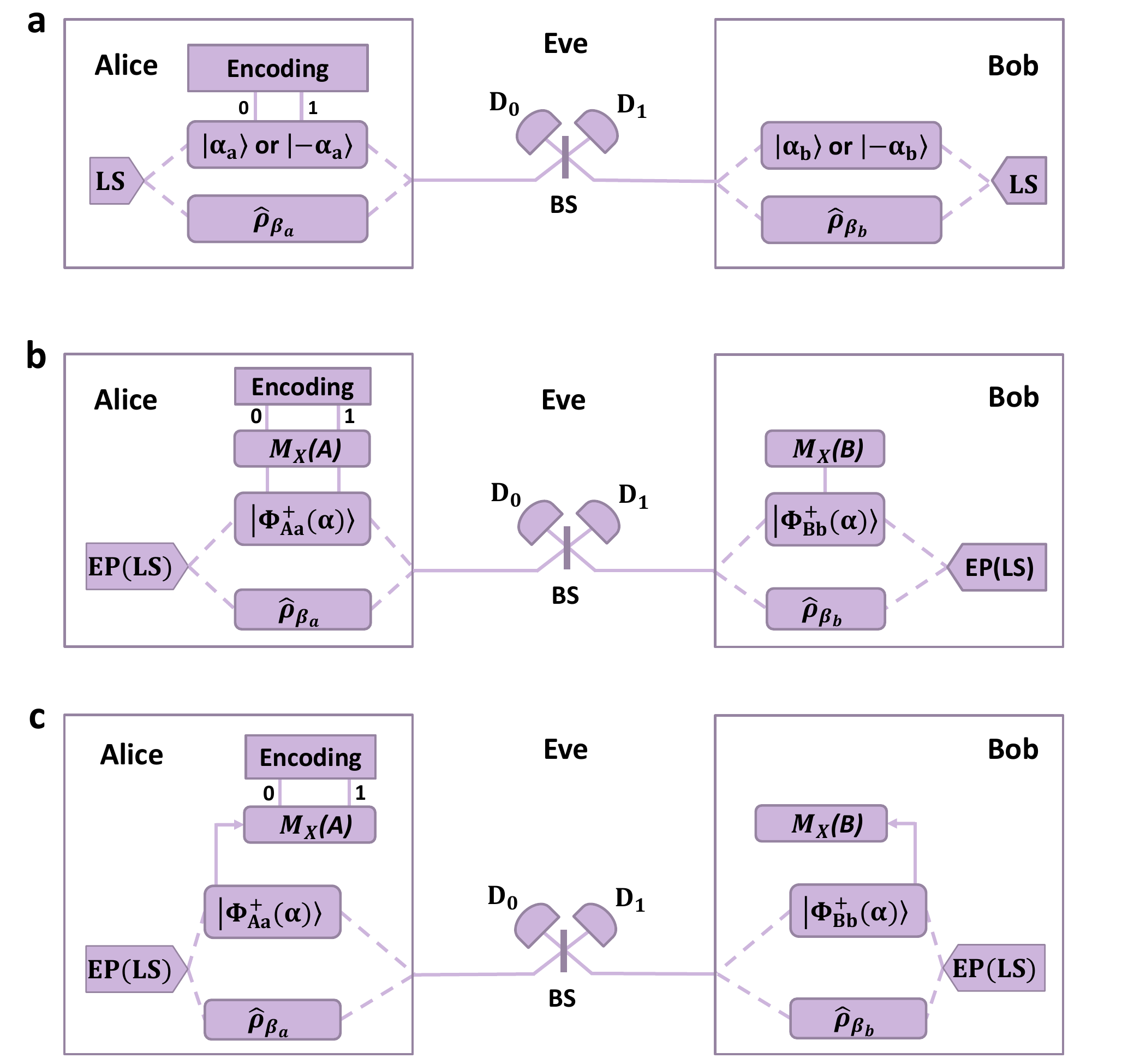}
     \caption{\label{fig:security}\textbf{Transmitters used for security analysis of OPI-QSDC.} LS: laser source; EP: entangled-photon source; BS, 50:50 beam splitter; $\rm D_0$, $\rm D_1$: single-photon detector; $\widehat{\rho }_{\beta_a}(\widehat{\rho }_{\beta_b})$: different amplitude and phase-randomized coherent states of Alice (Bob); $M_X( \cdot  )$: X-basis measurement. \textbf{a.} OPI-QSDC protocol. \textbf{b.} Equivalent view of OPI-QSDC protocol. \textbf{c.} Entanglement-based OPI-QSDC protocol.
     }
\end{figure}

In this section, we analyze the security of the OPI-QSDC protocol by analyzing the equivalent entanglement-based protocol. We first introduce the equivalence transformation process of our protocol, as shown in Fig.~\ref{fig:security}.

For convenience, we briefly show the process of the OPI-QSDC protocol in Fig.~\ref{fig:security}a. Charlie is not credible, so we replaced him with Eve. Alice and Bob use a coherent source, and randomly select one of two modes to send signals to Eve. In coding modes, the encoding of Alice includes all operations in \textbf{step 1}. Alice deterministically sends state $|\alpha\rangle_a$ or $|-\alpha\rangle_a$ to Eve depending on the encoding codes, while Bob randomly sends it. In multi-intensity modes, Alice and Bob send different amplitude and phase-randomized coherent states.

To randomly transmit bits 0 and 1, Bob can prepares an entangled state $|\Phi^+_{Bb}(\alpha)\rangle=(|+\rangle_{B}|\alpha\rangle_{b}+|-\rangle_{B}|-\alpha\rangle_{b})/\sqrt{2}$. Here, we use the subscript $B$ to represent the local qubit retained by Bob, and use the subscript $b$ to represent the photon to be sent to Eve. Then Bob uses the X basis to measure the local qubit. Similarly, Alice can prepare an entangled state $|\Phi^+_{Aa}(\alpha)\rangle$ and measures the local qubit in X basis to transmit bits 0 and 1. Without loss of generality, we assume Alice can make a deterministic measurement in the X basis based on the precoding result. For example, the precoded bit is 0, the measurement result will be $|+\rangle_A$, which makes Fig.~\ref{fig:security}a become Fig.~\ref{fig:security}b.

We can also consider a entanglement-based version of this protocol, namely the measurement of Alice and Bob can be delayed after Eve's announcement of the successful event. Therefore, we get Fig.~\ref{fig:security}c, which is our entanglement-based protocol. This measurement operation commutes with all the operations performed in the other steps. Hence this entanglement-based protocol is mathematically equivalent to the OPI-QSDC protocol. We will prove its security, thus proving the security of our OPI-QSDC protocol. The detailed steps of the equivalent entanglement-based protocol are as follows.

\textbf{Step 1: encoding.} Same as OPI-QSDC protocol.

\textbf{Step 2': mode preparation.} Alice and Bob randomly select the coding mode with a probability of $1-p$ and the multi-intensity mode with a probability of $p$, where $0<p<<1$.

\setlength{\parindent}{2em}Coding mode: Alice prepares an entangled state $|\Phi^+_{Aa}(\alpha)\rangle=(|+\rangle_{A}|\alpha\rangle_{a}+|-\rangle_{A}|-\alpha\rangle_{a})/\sqrt{2}$. She sends the photon $a$ to Eve, and retains the local photon $A$. Similarly, Bob prepares an entangled state  $|\Phi^+_{Bb}(\alpha)\rangle$ and sends the photon $b$ to Eve.

\setlength{\parindent}{2em}Multi-intensity mode: Same as OPI-QSDC protocol.

\textbf{Step 3': measurement.} Charlie measures the pulses sent by Alice and Bob using single-photon-interferometer and publishes measurement results. Alice and Bob discard the no-click events and two-click events, and retain the one-click events, namely $m_C\oplus m_D=1$. 
Then, if Alice (Bob) sent the coding mode, she (he) measures the retained local photon A (B) using the X basis. Here, we assume Alice can make a deterministic measurement based on the precoding result.

\textbf{Step 4: mode matching.} Same as OPI-QSDC protocol.

\textbf{Step 5: parameter estimation.} Same as OPI-QSDC protocol.

\textbf{Step 6: decoding.} Same as OPI-QSDC protocol.

We use ${|\Psi ^{\pm}\rangle,|\Phi^{\pm}\rangle}$ to mark Bell states, where
\begin{equation}
\begin{aligned}
 |\Psi^{-}_{AB}\rangle&=(|-+\rangle_{AB}-|+-\rangle_{AB})/\sqrt{2}\\
 &=(|01\rangle_{AB}-|10\rangle_{AB})/\sqrt{2},\\
 |\Psi^{+}_{AB}\rangle&=(|++\rangle_{AB}-|--\rangle_{AB})/\sqrt{2}\\
 &=(|01\rangle_{AB}+|10\rangle_{AB})/\sqrt{2},\\
 |\Phi^{-}_{AB}\rangle&=(|+-\rangle_{AB}+|-+\rangle_{AB})/\sqrt{2}\\
 &=(|00\rangle_{AB}-|11\rangle_{AB})/\sqrt{2},\\
 |\Phi^{+}_{AB}\rangle&=(|++\rangle_{AB}+|--\rangle_{AB})/\sqrt{2}\\
 &=(|00\rangle_{AB}+|11\rangle_{AB})/\sqrt{2}.\\
\end{aligned}
\end{equation}
We define $|\pm \rangle_{A(B)} \equiv (|0 \rangle_{A(B)}\pm|1\rangle_{A(B)})/\sqrt{2}$, where $|0\rangle_{A(B)}$ represents the vacuum state, and $|1\rangle_{A(B)}$ represents the single-photon state for optical pulse A (B). A coherent state $|\alpha \rangle$ is defined as 
\begin{equation}
  \label{coherent}
  \begin{aligned}
    |\alpha \rangle\equiv e^{-\frac{\left\lvert \alpha \right\rvert^2}{2} } \sum_{n = 0}^{\infty}\frac{(\alpha)^n}{\sqrt{n!} }|n\rangle.
  \end{aligned}
  \end{equation}
And We use the following transformation matrix corresponding to a 50:50 beam splitter:
\begin{equation}
  \label{BS}
\left(
  \begin{array}{c}
    c^\dagger \\ 
    d^\dagger
  \end{array}
\right)
=\frac{1}{\sqrt{2} } 
\left(
  \begin{array}{cc}
    1 & 1 \\
    1 & -1 
  \end{array}
\right)
\left(
  \begin{array}{c}
    a^\dagger \\
    b^\dagger 
  \end{array}
\right),
\end{equation}
where $a^\dagger$, $b^\dagger$ represents input optical pulse of a, b, repectively, and $c^\dagger$, $d^\dagger$ represents output optical pulse of C, D, repectively.

Next, we analyze the eavesdropping process and calculate $I(A:E)$. We first focus on the joint state $\rho_{AB}$ shared by Alice and Bob in the \textbf{Step 3'} after Charlie's measurement. If $\rho_{AB}$ is a pure state, it can be considered that there is no information leakage. Without loss of generality, we can assume that Alice and Bob currently share the entangled state $|\Psi^{+}_{AB}\rangle$. If eavesdropping occurs, $\rho_{AB}$ will become a mixed state. To facilitate the calculation, we can assume that Alice and Bob use some of the same unitary operations to change $\rho_{AB}$ into a diagonal matrix~\cite{wu2019security}, then
\begin{equation}
  \begin{aligned}
    \rho_{AB}=&\lambda_1|\Psi^{-}_{AB}\rangle \langle \Psi^{-}_{AB}|+\lambda_2|\Psi^{+}_{AB}\rangle \langle \Psi^{+}_{AB}|\\
    &+\lambda_3|\Phi^{-}_{AB}\rangle \langle \Phi^{-}_{AB}|+\lambda_4|\Phi^{+}_{AB}\rangle \langle \Phi^{+}_{AB}|.\\
  \end{aligned}
\end{equation}
The parameters satisfies $\sum_{i=1}^{4} \lambda_i=1$. We assume that Eve uses an auxiliary system $|E\rangle$ for eavesdropping and performs a collective attack, which can be extended to coherent attacks through the quantum de Finetti theorem~\cite{renner2007symmetry}. Then, a purification state $|\phi _{ABE}\rangle$ of $\rho_{AB}$ and $|E\rangle$ can be written as
\begin{equation}
  |\phi _{ABE}\rangle=\sqrt{\lambda _1}|\Psi^{-}_{AB}\rangle|E_1\rangle+\sqrt{\lambda _2}|\Psi^{+}_{AB}\rangle|E_2\rangle+\sqrt{\lambda _3}|\Phi^{-}_{AB}\rangle|E_3\rangle+\sqrt{\lambda _4}|\Phi^{+}_{AB}\rangle|E_4\rangle,
\end{equation}
where $|E_1\rangle$, $|E_2\rangle$, $|E_3\rangle$ and $|E_4\rangle$ are the orthogonal states of Eve's auxiliary system $|E\rangle$. If Alice and Bob do not share a Bell state $|\Psi^{+}_{AB}\rangle$, but instead share a Bell state $|\Phi^{+}_{AB}\rangle$ or $|\Phi^{-}_{AB}\rangle$, and they perform measurements in the Z basis, which will lead to a phase-error rate. Thus, the phase-error rate $E_u^Z$ is given by
\begin{equation}
  E_u^Z= \lambda_3+\lambda_4.
\end{equation} 
Eve can intercepts the qubits from Alice and acquires system AE:
\begin{equation}
  \rho _{AE}=Tr_B(|\phi _{ABE}\rangle\langle\phi _{ABE}| )=\frac{1}{2}(|\varphi  _1\rangle\langle \varphi _1|+|\varphi  _{2}\rangle\langle\varphi _{2}|),
\end{equation}
where
\begin{equation}
  \begin{aligned}
    |\varphi  _1\rangle\equiv &|-_A\rangle(\sqrt{\lambda_1}|E_1\rangle+\sqrt{\lambda_3} |E_3\rangle)+|+_A\rangle(\sqrt{\lambda_2}|E_2\rangle+\sqrt{\lambda_4} |E_4\rangle),\\
    |\varphi  _2\rangle\equiv &|-_A\rangle(-\sqrt{\lambda_2}|E_2\rangle+\sqrt{\lambda_4} |E_4\rangle)+|+_A\rangle(-\sqrt{\lambda_1}|E_1\rangle+\sqrt{\lambda_3} |E_3\rangle).
  \end{aligned}
\end{equation}
Then, Alice measures her pulse in X basis, and obtains the bit value $V_A$. The states of Eve become
\begin{equation}
  \begin{aligned}
    \rho ^{V_A=+}_E=&Tr_A(|+_A\rangle\langle +_A| \rho _{AE}|+_A\rangle\langle+ _A|)\\
    =&|\psi _1\rangle\langle \psi _1|+|\psi _2\rangle\langle \psi _2|,\\
    \rho ^{V_A=-}_E=&Tr_A(|-_A\rangle\langle -_A| \rho _{AE}|-_A\rangle\langle- _A|)\\
    =&|\psi _3\rangle\langle \psi _3|+|\psi _4\rangle\langle \psi _4|,
  \end{aligned}
\end{equation}
where
\begin{equation}
  \begin{aligned}
   |\psi _1\rangle\equiv &\sqrt{\lambda_2}|E_2\rangle+\sqrt{\lambda_4} |E_4\rangle,\\
   |\psi _2\rangle\equiv &-\sqrt{\lambda_1}|E_1\rangle+\sqrt{\lambda_3} |E_3\rangle,\\
   |\psi _3\rangle\equiv &\sqrt{\lambda_1}|E_1\rangle+\sqrt{\lambda_3} |E_3\rangle,\\
   |\psi _4\rangle\equiv &-\sqrt{\lambda_2}|E_2\rangle+\sqrt{\lambda_4} |E_4\rangle.
  \end{aligned}
\end{equation}
We get the mutual information $I(A:E)$ of the joint system $AE$
\begin{equation}
  \begin{aligned}
   I(A:E)= &S(\sum_{k}p_k \rho ^{V_A=k}_E )-\sum_{k}p_kS(\rho ^{V_A=k}_E),\\
  \end{aligned}
\end{equation}
where $S(\cdot )$ is the von Neumann entropy, and $p_k$ is the distribution of $\rho ^{V_A=k}_E$. Supporting that Alice sends the same number of bits 0 and 1, so $p_k=\frac{1}{2} $. The upper bound on $I(A:E)$ is
\begin{equation}
  \begin{aligned}
  &I(A:E)\\
  =&H(\lambda_1,\lambda_2,\lambda_3,\lambda_4)-h(\lambda_1+\lambda_3)\\
  =&(\lambda_1+\lambda_3)h(\frac{\lambda_1}{\lambda_1+\lambda_3}) +(\lambda_2+\lambda_4)h(\frac{\lambda_2}{\lambda_2+\lambda_4})\\
  \leq &h(\lambda_3+\lambda_4)\\
  =&h(E_u^{Z}),
  \end{aligned}
\end{equation}
where $H(\lambda_1,\lambda_2,\lambda_3,\lambda_4)\equiv -\lambda_1\log_2\lambda_1-\lambda_2\log_2\lambda_2-\lambda_3\log_2\lambda_3-\lambda_4\log_2\lambda_4$. Therefore, if we estimate the phase-error rate $E_u^Z$, we can calculate the amount of information leakage. 

Next, we analyze the process of our protocol to estimate the phase-error rate $E_u^Z$. In the \textbf{Step 2'}, the beam splitter transforms the joint state $|\phi_{AaBb}\rangle=|\Phi^+_{Aa}(\alpha)\rangle\otimes |\Phi ^+_{Bb}(\alpha)\rangle$ into the following state
\begin{equation}
  \begin{aligned}
    \label{ABCD1}
  |\phi_{ABCD}\rangle=&\frac{1}{2}(|++\rangle_{AB}|\sqrt{2}\alpha,0\rangle_{CD}+|+-\rangle_{AB}|0,\sqrt{2}\alpha\rangle_{CD} \\
  &+|-+\rangle_{AB}|0,-\sqrt{2}\alpha\rangle_{CD}+|--\rangle_{AB}|-\sqrt{2}\alpha ,0\rangle_{CD}).\\
  \end{aligned}
\end{equation}
Here we assume $|\alpha\rangle_a=|\alpha\rangle_b=|\alpha\rangle$. And we have
\begin{equation}
  \begin{aligned}
  |\sqrt{2}\alpha\rangle=&e^{-\left\lvert \alpha \right\rvert^2 } \sum_{n = 0}^{\infty}\frac{(\sqrt{2}\alpha)^n}{\sqrt{n!} }|n\rangle\\
   =&\sum_{n = 0}^{\infty}C_{2n}|2n\rangle+\sum_{n = 0}^{\infty}C_{2n+1}|2n+1\rangle  ,\\
   |-\sqrt{2}\alpha\rangle=&e^{-\left\lvert \alpha \right\rvert^2 } \sum_{n = 0}^{\infty}\frac{(-\sqrt{2}\alpha)^n}{\sqrt{n!} }|n\rangle\\
   =&\sum_{n = 0}^{\infty}C_{2n}|2n\rangle-\sum_{n = 0}^{\infty}C_{2n+1}|2n+1\rangle  ,\\
  \end{aligned}
\end{equation}
where
\begin{equation}
  \begin{aligned}
  C_{2n}\equiv &e^{-\left\lvert \alpha \right\rvert^2 } \frac{(\sqrt{2}\alpha)^{2n}}{\sqrt{(2n)!} } ,\\
  C_{2n+1}\equiv &e^{-\left\lvert \alpha \right\rvert^2 } \frac{(\sqrt{2}\alpha)^{2n+1}}{\sqrt{(2n+1)!} } .
  \end{aligned}
\end{equation}
After arraging the Eq.~(\ref{ABCD1}), we obtain
\begin{equation}
  \begin{aligned}
    \label{ABCD2}
  |\phi_{ABCD}\rangle=&\frac{1}{2}[-|\Psi^-_{AB} \rangle|0\rangle_C(\sum_{n = 0}^{\infty}C_{2n+1}|2n+1\rangle_D)+|\Psi^+_{AB} \rangle(\sum_{n = 0}^{\infty}C_{2n+1}|2n+1\rangle_C)|0\rangle_D \\
  &+|\Phi^-_{AB} \rangle|0\rangle_C(\sum_{n = 0}^{\infty}C_{2n}|2n\rangle_D)+|\Phi^+_{AB} \rangle(\sum_{n = 0}^{\infty}C_{2n}|2n\rangle_C)|0\rangle_D].
  \end{aligned}
\end{equation}
We consider the case where only detector $\rm D_0$ click and define the measurement operator of the detectors at this moment as $\hat{M}_C$, where
\begin{equation}
  \hat{M}_C\equiv \sum_{n} \sqrt{Y^C_n}|n0\rangle _{CD} \langle n0|,
\end{equation}
where $Y^C_{n}=\langle n0_{CD}| \hat{M}_C^\dagger \hat{M}_C|n0_{CD}\rangle$ is the yield of the $n$-photon state, i.e., the conditional probability of only detector $\rm D_0$ click when there are $n$ photons in the channel. Tracing out system C and D from system ABCD after Charlie's measurement, we get the joint state $\rho_{AB}^C$ of Alice and Bob in the \textbf{Step 3'}, which is
\begin{equation}
  \begin{aligned}
    \rho_{AB}^C&=Tr_{CD}(\frac{\hat{M}_C}{\sqrt{Q_u^C} }  |\phi_{ABCD}\rangle \langle\phi_{ABCD}| \frac{\hat{M}_C^\dagger}{\sqrt{Q_u^C} })\\
    &=\frac{1}{Q^C_u}(\sum_{n = 0}^{\infty} \sqrt{|C_{2n+1}|^2 Y^C_{2n+1}})^2 |\Psi^+_{AB} \rangle \langle\Psi^+_{AB}|+\frac{1}{Q^C_u}(\sum_{n = 0}^{\infty} \sqrt{|C_{2n}|^2 Y^C_{2n}})^2 |\Phi^+_{AB} \rangle \langle\Phi^+_{AB}|.
  \end{aligned}
\end{equation}
where $Q^C_u=\langle \phi_{ABCD}| \hat{M}_C^\dagger \hat{M}_C|\phi_{ABCD}\rangle$ is the total gain, i.e., the conditional probability of only detector $\rm D_0$ click when Alice and Bob send pulses with the intensity of $u$. Therefore, we have
\begin{equation}
  \begin{aligned}
    E_u^{Z,C}&=\lambda_3+\lambda_4\\
    &=\frac{1}{Q^C_u}(\sum_{n = 0}^{\infty} \sqrt{|C_{2n}|^2 Y^C_{2n}})^2.
  \end{aligned}
\end{equation}
Hence the achievable rate is
\begin{equation}
  \label{rc}
  \begin{aligned}
   R^C=&I^C(A:B)-I^C(A:E)\\
   \geq&  q\cdot Q^C_u\cdot [1-fh(E^{X,C}_u)-h(E^{Z,C}_u)],
  \end{aligned}
\end{equation}
where $q=1-2p(1-p)$ is the mode matching rate, and $u=\left\lvert \alpha\right\rvert^2 $ is the mean photon number or the light intensity. $f\geq 1$ is an inefficiency function for forward coding. We skip the discussion for only $\rm D_1$ click, however, it holds in a similar manner, which is
\begin{equation}
  \begin{aligned}
   R^D\geq q\cdot Q^D_u\cdot [1-fh(E^{X,D}_u)-h(E^{Z,D}_u)].
  \end{aligned}
\end{equation}
Finally, the total secrecy rate formula is given by
\begin{equation}
  \begin{aligned}
   R=& \max\{R^C,0\}+\max\{R^D,0\}.
  \end{aligned}
\end{equation}

\section{Derivation details of parameterized OPI-QSDC protocol}
\label{app:Performance}
To determine the performance of OPI-QSDC protocol, we use the channel model in~\cite{ma2018phase}, which transfer $a^\dagger$, $b^\dagger$ according to
\begin{equation}
  \label{ab}
  \begin{aligned}
    a^\dagger\rightarrow&\sqrt{\eta }a^\dagger+\sqrt{1-\eta}a^\dagger_\bot ,\\
    b^\dagger\rightarrow&\sqrt{\eta }b^\dagger+\sqrt{1-\eta}b^\dagger_\bot,
  \end{aligned}
\end{equation}
where $\eta=\eta_d\sqrt{\eta_c}$ is the transmittance, and $\eta_d$, $\eta_c$ are the detection efficiencies and the channel losses function, respectively. The channel losses function is $\eta_c=10^\wedge (-\zeta d/10)$, where $\zeta$ represents the attenuation coefficient, and $d$ represents the transmission distance. The transformation matrix of the beam splitter is the same as Eq.~(\ref{BS}), which leads to the the transformations of the coherent state as
\begin{equation}
  \label{2coherent}
  \begin{aligned}
    &|\sqrt{u} e^{i\phi _a}\rangle_a |\sqrt{u} e^{i\phi _b}\rangle_b\rightarrow |\sqrt{\frac{\eta u}{2} }(e^{i\phi _a}+e^{i\phi _b})\rangle_C |\sqrt{\frac{\eta u}{2} }(e^{i\phi _a}-e^{i\phi _b})\rangle_D.
  \end{aligned}
\end{equation}
Here $\phi _a$, $\phi _b \in \{0,\pi \}$.

We consider the case where only detector $\rm D_0$ click, i.e. $\phi _a=\phi _b $. To model the phase mismatch between Alice and Bob's signals arriving at Charlie, we introduce a parameter $\delta \equiv \phi _b-\phi _a$. We first consider the case when there is only one photon in the channel $(a^\dagger+e^{i\delta}b^\dagger)|00\rangle_{ab}$. Hence the detection probabilities of Charlie are
\begin{equation}
  \begin{aligned}
    &q^0_1=1-\eta,\\
    &q^C_1=\eta \cos^2(\frac{\delta }{2}),\\
    &q^D_1=\eta \sin^2(\frac{\delta }{2}),\\
    &q^{C,D}_1=0,
  \end{aligned}
\end{equation}
where $q^0_1$, $q^C_1$, $q^D_1$ and $q^{C,D}_1$ are the probabilities for no click, $\rm D_0$ click, $\rm D_1$ click an double click, respectively. 

Then we consider the case when there is $n$ photons in the channel $(a^\dagger+e^{i\delta}b^\dagger)^n|00\rangle_{ab}$, for which we can regard as $n$ identical and independent click events of the one-photon case~\cite{ma2018phase}. We take into the effects of the detector dark counts $p_d$, thus the detection probabilities of the detector $\rm D_0$ and $\rm D_1$ are
\begin{equation}
  \begin{aligned}
    q^C_n=&p_d(1-p_d)(1-\eta)^n+(1-p_d)\{[1-\eta+\eta \cos^2(\frac{\delta }{2})]^n-(1-\eta)^n\},\\
    q^D_n=&p_d(1-p_d)(1-\eta)^n+(1-p_d)\{[1-\eta+\eta \sin^2(\frac{\delta }{2})]^n-(1-\eta)^n\}.\\
  \end{aligned}
\end{equation}
Based on the above equation, the yield $Y^C_n$ of the $n$-photon state is given by
\begin{equation}
  \label{yield}
  \begin{aligned}
   Y^C_n=&q^C_n+q^D_n\\
   =&1-(1-2p_d)(1-\eta)^n.
  \end{aligned}
\end{equation}
Here, we take the first order of a small $\delta$. 

Finally, we consider the case of both Alice and Bob to send the coherent state
\begin{equation}
  \label{2coherent_2}
  \begin{aligned}
    &|\sqrt{u} e^{i\phi _a}\rangle_a |\sqrt{u} e^{i\phi _b}\rangle_b\\
    =&e^{-u}e^{\sqrt{u}e^{i\phi _a}(a^\dagger+e^{i\delta }b^\dagger)}|00\rangle_{ab} \\
    =&e^{-u}\sum_{n} \frac{(\sqrt{u}e^{i\phi _a} )^n}{\sqrt{n!}}(a^\dagger+e^{i\delta }b^\dagger)^n |00\rangle_{ab}.
  \end{aligned}
\end{equation}
 In line with Eqs.~(\ref{2coherent}), (\ref{2coherent_2}), Charlie measures the coherent state $|\pm \sqrt{\frac{\eta u}{2}}(1+e^{i\delta})\rangle$ entering the detector $\rm D_0$. When it collapses to the fock state of $|0\rangle$, there will be no click. Thus, the detection probabilities are
\begin{equation}
  \begin{aligned}
    \overline{q^C_u} =&(1-p_d)\exp[-2\eta u\cos^2(\frac{\delta}{2} )],\\
    q^C_u =&1-\overline{q^C_u},\\
    \overline{q^D_u} =&(1-p_d)\exp[-2\eta u\sin^2(\frac{\delta}{2} )],\\
    q^D_u =&1-\overline{q^D_u},
  \end{aligned}
\end{equation}
where $\overline{q^C_u}$ and $\overline{q^D_u}$ are the probabilities of the no click. Therefore, the total gain $Q^C_u$ is given by
\begin{equation}
  \label{gain}
  \begin{aligned}
    Q^C_u=&\overline{q^C_u}q^D_u+q^C_u\overline{q^D_u}\\
    =&1-e^{-2\eta u}+2p_de^{-2\eta u}.
  \end{aligned}
\end{equation}

The X-basis QBER are caused by dark counts,
\begin{equation}
  \label{ex}
  \begin{aligned}
    E^{X,C}_u=&\frac{\overline{q^C_u}q^D_u}{Q^C_u} \\
    =&\frac{e^{-2\eta u}}{Q^C_u} [p_d+2\eta u\sin^2(\frac{\delta}{2} )].
  \end{aligned}
\end{equation}
And the even-photon signals in the channel will cause the Z-basis error, i.e. phase error. In practical implementation, the multi-intensity mode can be employed to estimate the yields $Y^C_{2n}$ of even-photon fock states. Thus the phase randomization is performed in the Z basis, which makes the photon number distribution of the coherent state follow a Poisson distribution,
\begin{equation}
  \begin{aligned}
    P(n)=e^{-2u}\frac{(2u)^n}{n!}.
  \end{aligned}
\end{equation}
Here, we note that $C_{2n}=\sqrt{P(2n)} $ and we can assume that there is no eavesdropper in the channel which allows us to utilize Eq.~(\ref{yield}) to estimate the Z-basis error,
\begin{equation}
  \label{ez}
  \begin{aligned}
    Q^C_u E^{Z,C}_u=&(\sum_{n = 0}^{\infty} \sqrt{ P(2n)Y^C_{2n}})^2.
  \end{aligned}
\end{equation}

Therefore, when $m_C=1$, $m_D=0$, the secrecy rate $R^C$ is derivated by combining Eqs.~(\ref{rc}), (\ref{gain}), (\ref{ex}), (\ref{ez}). The secrecy rate $R^D$ is same as $R^C$ by repeating the above steps, thus the total secrecy rate $R$ is derivated. 

\section{Simulation formulas for DL04 and MDI-QSDC protocol}
\label{app:DL04andMDI}
The secrecy rate of DL04 protocol with INCUM is given by~\cite{long2021drastic}
\begin{equation}
C_s^M\geq   \max\{Q_B[1-h(e)-h(\varepsilon _x+\varepsilon _z)],0\}.
\end{equation}
In the simulation, the gain of Alice $Q_A$, Bob $Q_B$ and QBER $e$ is given by~\cite{long2021drastic}
\begin{equation}
  \begin{aligned}
    Q_A&=\eta_d \eta_c+p_d,\\
    Q_B&=\eta_d \eta_c^2+p_d,\\
    eQ_B&=e_d\eta_d \eta_c^2+e_0p_d,
  \end{aligned}
\end{equation}
where $e_d=1.3\%$ is the intrinisic detector error rate, and $e_0=1/2$ is the background error rate. For simplicity, we assume that the detection error rate $\varepsilon _x=\varepsilon _z=\varepsilon$, and it is given by
\begin{equation}
  \varepsilon Q_A=e_d\eta_d \eta_c+e_0p_d.
\end{equation}

The secrecy rate of MDI-QSDC protocol with INCUM is given by~\cite{li2022single}
\begin{equation}
  C_s=Q[1-h(e)-h(\epsilon_y)],
\end{equation}
where $Q$ is the signal gain of Bob for message decoding, while $e$ and $\epsilon_y$ represent the QBER and the detection error rate, respectively. And the gain $Q$ is given by~\cite{Niu2020Research,li2022single}
\begin{equation}
  \begin{aligned}
    Q&=q_{C1}q_{C2},\\
    q_{C1}&=\frac{\eta_c}{3} (G_X+G_Z),\\
    G_X&=G_Y=\mathcal{P}_{12}^{-+}+\mathcal{P}_{14}^{-+}+\mathcal{P}_{12}^{++}+\mathcal{P}_{14}^{++},\\
    G_Z&=2(\mathcal{P}_{ij}^{HV}+\mathcal{P}_{ij}^{HH}),\\
    q_{C2}&=\eta_c+(1-\eta_c)p_d.
  \end{aligned}
\end{equation}
Here,
\begin{equation}
  \mathcal{P}_{ij}^{HV}=\mathcal{P}_{ij}^{VH}=(1-\eta_c)^2p_d^2(1-p_d)^2+(1-\eta_c)\eta_cp_d(1-p_d)^2+\frac{1}{4}\eta_c^2(1-p_d)^2.
  \end{equation}
  \begin{equation}
  \mathcal{P}_{ij}^{HH}=\mathcal{P}_{ij}^{VV}=(1-\eta_c)^2p_d^2(1-p_d)^2+(1-\eta_c)\eta_cp_d(1-p_d)^2+\frac{1}{2}\eta_c^2p_d(1-p_d)^2.
  \end{equation}
  \begin{equation}
  \mathcal{P}_{12}^{-+}=\mathcal{P}_{12}^{+-}=\mathcal{P}_{34}^{-+}=\mathcal{P}_{34}^{+-}=(1-\eta_c)^2p_d^2(1-p_d)^2+(1-\eta_c)\eta_cp_d(1-p_d)^2+\frac{1}{4}\eta_c^2p_d(1-p_d)^2.
  \end{equation}
  \begin{equation}
  \mathcal{P}_{14}^{-+}=\mathcal{P}_{14}^{+-}=\mathcal{P}_{23}^{-+}=\mathcal{P}_{23}^{+-}=(1-\eta_c)^2p_d^2(1-p_d)^2+(1-\eta_c)\eta_cp_d(1-p_d)^2+\frac{1}{4}\eta_c^2(p_d+1)(1-p_d)^2.
  \end{equation}
  \begin{equation}
  \mathcal{P}_{12}^{++}=\mathcal{P}_{34}^{++}=\mathcal{P}_{12}^{--}=\mathcal{P}_{34}^{--}=(1-\eta_c)^2p_d^2(1-p_d)^2+(1-\eta_c)\eta_cp_d(1-p_d)^2+\frac{1}{4}\eta_c^2(p_d+1)(1-p_d)^2.
  \end{equation}
  \begin{equation}
  \mathcal{P}_{14}^{++}=\mathcal{P}_{23}^{++}=\mathcal{P}_{14}^{--}=\mathcal{P}_{23}^{--}=(1-\eta_c)^2p_d^2(1-p_d)^2+(1-\eta_c)\eta_cp_d(1-p_d)^2+\frac{1}{4}\eta_c^2p_d(1-p_d)^2.
  \end{equation}
the QBER and the detection error rate is given by~\cite{Niu2020Research,li2022single}
\begin{equation}
  \begin{aligned}
  e=&\frac{e_0p_d+e_{\rm edt}\eta_c}{\eta_c+(1-\eta_c)p_d},\\
  \epsilon_y=&e_{d}(1-2\hat{\epsilon_y})+\hat{\epsilon_y},\\
  \hat{\epsilon_y}=&\frac{\sum\limits_{(i, j)\in\{(1,4), (2,3)\}}(\mathcal{P}_{ij}^{++}+\mathcal{P}_{ij}^{--})+\sum\limits_{(i, j)\in\{(1,2), (3,4)\}}(\mathcal{P}_{ij}^{+- }+\mathcal{P}_{ij}^{-+})}{4G_Y}.\\
  \end{aligned}
\end{equation}

\section*{References} 
\bibliographystyle{iopart-num}
\bibliography{opiqsdc.bib}

\end{document}